\documentclass[aps,pra,twocolumn,showpacs]{revtex4-1}
\usepackage{epsfig}
\usepackage{bm}

\newcommand{\be}{\begin{eqnarray}}
\newcommand{\ee}{\end{eqnarray}}
\newcommand{\ba}{\begin{array}}
\newcommand{\ea}{\end{array}}
\newcommand{\bmat}{\left(\begin{array}}
\newcommand{\emat}{\end{array}\right)}
\newcommand{\no}{\nonumber}
\newcommand{\Tr}{\mbox{Tr}\,}
\newcommand{\diff}{\mathrm d}
\newcommand{\e}{\mathrm e}

\begin{document}
\title{Unitary deformations of counterdiabatic driving}
\author{Kazutaka Takahashi}
\affiliation{Department of Physics, Tokyo Institute of Technology, 
Tokyo 152-8551, Japan}

\date{\today}

\begin{abstract}
We study a deformation of the counterdiabatic-driving Hamiltonian 
as a systematic strategy for an adiabatic control of quantum states.
Using a unitary transformation, we design
a convenient form of the driver Hamiltonian.
We apply the method to a particle in a confining potential and 
discrete systems to find explicit forms of the Hamiltonian 
and discuss the general properties.
The method is derived by using the quantum brachistochrone equation,
which shows the existence of a nontrivial dynamical invariant
in the deformed system.
\end{abstract}
\maketitle

\section{Introduction}

The method ``shortcuts to adiabaticity'' has attracted 
much attention recently.
A state evolution by the adiabatic process is 
accelerated by applying the counterdiabatic Hamiltonian
and many applications have been 
discussed~\cite{DR1,DR2,Berry,CRSCGM,STA}.
From a more fundamental perspective, the method is understood 
as an optimized protocol~\cite{CHKO,Takahashi}.
Furthermore, it has been shown to be useful
in real experiments~\cite{SSVL,SSCVL,Betal,Zetal}.

Some technical problems arise when we apply the method
for a given adiabatic Hamiltonian.
First, we need all eigenstates of 
the Hamiltonian to construct the counterdiabatic term.
It is generally a difficult task except the case where 
the Hamiltonian takes a simple form in a small Hilbert space.
Second, even if the counterdiabatic term is found,
it involves operators which are difficult to implement in the laboratory.

These problems can be circumvented when we use  
the invariant-based engineering which is known to be 
equivalent essentially to the counterdiabatic driving~\cite{CTM}.
To find an ideal control, we first specify the form of the invariant 
instead of preparing the adiabatic Hamiltonian. 
Thus, the initial setting is different in both methods 
and the availability depends on the problem to consider.
Therefore, it will be useful if we have other choices.
As possible solutions of the problems, 
various methods have been discussed intensively: 
approximations of the counterdiabatic term by 
a simple operator~\cite{dCRZ,OM}, 
unitary transformations of the state~\cite{ICTMR,MTCM}, 
modification of the formula using arbitrariness of 
the counterdiabatic term~\cite{Takahashi2}, and 
use of scale invariance~\cite{delCampo,dCB,DJdC}.

In this paper, we discuss a strategy for constructing a driver Hamiltonian.
Using a unitary transformation, we design the counterdiabatic term 
in a convenient form.
The same strategy was used in many works 
such as Refs.~\cite{ICTMR,MTCM,delCampo,DJdC}.
We go further to obtain a convenient form of the driver Hamiltonian.
There is some arbitrariness in choosing the driver Hamiltonian 
as discussed in Ref.~\cite{Takahashi2} if we control a specific state.
We exploit this arbitrariness to find the counterdiabatic potential.

We should also mention a similarity of our method to 
the fast-forward scaling~\cite{MN1,MN2,MN3,TMRM,MR,Takahashi3} 
where the state evolution is fast-forwarded by using the time scaling.
To find the fast-forward potential, 
a position-dependent phase is added to the wave function.
It is important to notice that the point of the method is 
not in the scaling but in the addition of the phase.
The phase is nothing but a unitary transformation, 
to be discussed more generally in the present paper.
Thus, the same idea is used by many works.
Our aim is to combine the ideas used in many contexts
and to clarify the general properties of the method.

The organization of the paper is as follows.
In Sec.~\ref{ud}, the basic idea of our method is described.
By applying the idea to a particle in a potential in Sec.~\ref{pot} and 
discrete systems in Sec.~\ref{discrete}, 
we discuss general properties of the method. 
In Sec.~\ref{qb}, we formulate the problem using 
the quantum brachistochrone equation 
to see the method from a different perspective.
Finally, the summary is given in Sec.~\ref{sum}.

\section{Unitary deformation}
\label{ud}

We describe our basic idea of the unitary deformation.
We start from the Schr\"odinger equation 
\be
 i\frac{\diff}{\diff t}|\psi(t)\rangle
 = \hat{H}(t)|\psi(t)\rangle. \label{Sch}
\ee
For a given time-dependent Hamiltonian $\hat{H}(t)$ and 
an initial condition of state $|\psi(0)\rangle$, 
$|\psi(t)\rangle$ is obtained by solving the equation.
In the method of counterdiabatic driving, the Hamiltonian is
divided into two parts: 
\be
 && \hat{H}(t)=\hat{H}_{\rm ad}(t)+\hat{H}_{\rm cd}(t), \\
 && \hat{H}_{\rm ad}(t)=\sum_n E_n(t)|n(t)\rangle\langle n(t)|, 
 \label{Had} \\
 && \hat{H}_{\rm cd}(t)=i\sum_{\stackrel{\scriptstyle m,n}{(m\ne  n)}}
 |m(t)\rangle\langle m(t)|\dot{n}(t)\rangle\langle n(t)|,
 \label{Hcd}
\ee
where $E_n(t)$ is an instantaneous eigenvalue of the adiabatic Hamiltonian
$\hat{H}_{\rm ad}(t)$ and $|n(t)\rangle$ is the corresponding eigenstate.
The dot denotes the time derivative 
as $|\dot{n}(t)\rangle=\frac{\diff}{\diff t}|n(t)\rangle$.
The solution of the Schr\"odinger equation is given by 
a linear combination of the adiabatic states of $\hat{H}_{\rm ad}(t)$ 
defined as 
\be
 |\psi_n(t)\rangle = \exp\left[-i\int^t_0\diff t'\,
 \left(E_n(t')-i\langle n(t')|\dot{n}(t')\right)\right]|n(t)\rangle. \no\\
\ee
This result is very convenient for the ideal control of the system 
since the deviation from the adiabatic state is prevented  
by the presence of the counterdiabatic term $\hat{H}_{\rm cd}(t)$.

The assumption here is that the form of $\hat{H}_{\rm cd}(t)$ becomes 
complicated and is hard to be realized.
Although the counterdiabatic term is specified as in Eq.~(\ref{Hcd}), 
its formal representation based on a spectral decomposition 
is not a useful one.
We introduce a unitary transformation: 
\be
 |\tilde{\psi}(t)\rangle = \hat{U}_\psi(t)|\psi(t)\rangle.
\ee
The Schr\"odinger equation takes the form 
\be
 && i\frac{\diff}{\diff t}|\tilde{\psi}(t)\rangle
 = \hat{H}_U(t)|\tilde{\psi}(t)\rangle, \\
 && \hat{H}_U(t)=\hat{U}_\psi(t)\hat{H}(t)\hat{U}_\psi^\dag(t)
 -i\hat{U}_\psi(t)\frac{\diff\hat{U}_\psi^\dag(t)}{\diff t}.
\ee
We want to find a convenient form of potential term $\hat{V}_\psi(t)$
such that the following relation is satisfied:
\be
 \hat{H}_U(t)|\tilde{\psi}(t)\rangle
 = (\hat{H}_{\rm ad}(t)+\hat{V}_\psi(t))|\tilde{\psi}(t)\rangle. 
\ee
We note that the equality 
$\hat{H}_U(t)=\hat{H}_{\rm ad}(t)+\hat{V}_\psi(t)$
is not necessarily satisfied.
With the potential term, 
the state follows an adiabatic passage which is different from 
the original one although 
the adiabatic Hamiltonian is the same in both the evolutions.
In realistic applications developing a specific initial state to 
a final one, we impose the condition that 
the unitary operator $\hat{U}_\psi(t)$ goes to zero at 
initial and final times.
We consider such conditions in the following examples.

There is of course a large amount of arbitrariness 
in the choice of the unitary transformation.
Our aim is to establish the general strategy to determine 
the driving potential $\hat{V}_\psi(t)$.
We note that the form of the potential $\hat{V}_\psi(t)$ crucially depends 
on the state $|\psi(t)\rangle$ to accelerate.
If we change the state, the form will be changed accordingly.
The original counterdiabatic driving is applied to any state.
By abandoning such a universal property, 
we can use the convenient acceleration potential.
However, we show in the following that 
the potential can be effectively independent of states in some special cases.

\section{Infinite Hilbert space}
\label{pot}

In this section, we study one-particle systems in a position-dependent potential.
The adiabatic Hamiltonian is given by
\be
 \hat{H}_{\rm ad}(t)=\frac{\hat{\bm{p}}^2}{2m}+U(\hat{\bm{r}},t).
\ee
A time dependence is present in the potential function $U$.

\subsection{One-dimensional system}

For one-dimensional bound states in a confining potential $U(\hat{x},t)$, 
the wave function is real 
and one can simplify the formula of the potential.
The adiabatic state corresponding to an eigenvalue $E_n(t)$ 
is represented as 
\be
 \psi_n(x,t)=\exp\left(-i\int_0^t \diff t'\, E_n(t')\right)\varphi_n(x,t),
 \label{psiad}
\ee
where $\varphi_n(x,t)$ is real.
We note that the additional phase 
due to the parallel transport is absent since 
$\langle\varphi_n(t)|\dot{\varphi}_n(t)\rangle=0$.

In this system, the counterdiabatic term 
depends not only on the position operator $\hat{x}$
but also on the momentum operator $\hat{p}$,
which is inconvenient in practical applications.
We consider the unitary deformation to control the system
by a local potential $V_n(\hat{x},t)$.
We define the new state 
\be
 \tilde{\psi}_n(x,t)=\e^{-i\phi_n(x,t)}\psi_n(x,t),
 \label{phase}
\ee
where $\phi_n(x,t)$ is a real function.
This new state satisfies the Schr\"odinger equation 
\be
 i\frac{\partial}{\partial t}\tilde{\psi}_n(x,t)
 = \left(-\frac{1}{2m}\frac{\partial^2}{\partial x^2}
 +U(x,t)+V_n(x,t)\right)\tilde{\psi}_n(x,t). \no\\
\ee
Using the properties that $\psi_n$ is factorized as Eq.~(\ref{psiad})
and $\varphi_n(x,t)$ is real, we obtain 
\be
 && \frac{\partial\phi_n}{\partial t}+E_n
 =\frac{1}{2m}\left(\frac{\partial\phi_n}{\partial x}\right)^2+U+V_n
 -\frac{1}{2m}\frac{\frac{\partial^2\varphi_n}{\partial x^2}}{\varphi_n}, \\
 && \frac{\partial\varphi_n^2}{\partial t}
 =\frac{1}{m}\frac{\partial}{\partial x}\left(
 \frac{\partial\phi_n}{\partial x}\varphi_n^2\right).
\ee
The first equation at the classical limit 
represents the Hamilton-Jacobi equation 
and the second one is the continuity equation.
We note that 
the role of the additional phase $\phi_n(x,t)$
is to induce a current with the probability density unchanged.
The current is exactly equal to zero in the original wave function, 
which is not appropriate to describe the present dynamical problem.

Since $\varphi_n(x,t)$ represents the eigenstate of the adiabatic
Hamiltonian, we have the relation
\be
 E_n(t)=-\frac{1}{2m}
 \frac{\frac{\partial^2\varphi_n(x,t)}{\partial x^2}}{\varphi_n(x,t)}+U(x,t).
\ee
Using this, we can write the potential as
\be
 V_n(x,t) = \frac{\partial\phi_n(x,t)}{\partial t}
 -\frac{1}{2m}\left(\frac{\partial\phi_n(x,t)}{\partial x}\right)^2.
 \label{Vn}
\ee
The phase $\phi_n(x,t)$ is determined from the continuity equation as 
\be
 \phi_n(x,t)
 =m\int^x_{x^*} \diff x_1\,\frac{\int_{x^*}^{x_1} \diff x_2\,
 \frac{\partial\rho_n(x_2,t)}{\partial t}}{\rho_n(x_1,t)},
 \label{phin}
\ee
where $x^*$ is a reference point and $\rho_n(x,t)=\varphi_n^2(x,t)$
represents the probability density.
Thus, once if we know the probability density of the adiabatic state,
we can calculate the counterdiabatic potential.
We note that it is not necessary to know the form of the counterdiabatic 
term to find the potential.
Therefore, this formula can be a general strategy 
to obtain the counterdiabatic driving.

The form of the potential depends on the adiabatic state
as it is labeled by the index $n$.
We show that it is independent of $n$ for several typical situations.
First, we consider the transport dynamics.
The potential is given by 
\be
 U(x,t)=U_0(x-x_0(t)), \label{Utr}
\ee
where $x_0(t)$ is a real function 
and describes the translation of the potential.
The probability density takes the form 
\be
 \rho_n(x,t)=f_n(x-x_0(t)).
\ee
In this case, the phase is obtained from the continuity equation as 
\be
 \phi_n(x,t)
 &=& -m\dot{x}_0(t)(x-x_0(t))+m\dot{x}_0(t)(x^*-x_0(t))
 \no\\ & &
 +m\dot{x}_0(t)f_n(x^*-x_0(t))
 \int^{x}_{x^*} \frac{\diff x_1}{f_n(x_1-x_0(t))}.
 \no\\ \label{phix0}
\ee
The corresponding potential is calculated from Eq.~(\ref{Vn}).
The result depends on the function $f_n$, which means that
the counterdiabatic potential crucially depends on the state to accelerate.
However, the last term in Eq.~(\ref{phix0}) can be neglected
if we set $f_n(x^*-x_0(t))=0$.
For bound states, we can set $x^*=\infty$ or $-\infty$.
In this case, the second term in Eq.~(\ref{phix0}) is divergent but 
it is an irrelevant energy shift and 
does not affect the adiabatic state.
We can use the phase 
\be
 \phi_n(x,t) = -m\dot{x}_0(t)(x-x_0(t)).
\ee
The potential is given by 
\be
 V_n(x,t) 
 &=& -m\ddot{x}_0(t)(x-x_0(t))+\frac{m}{2}\dot{x}_0^2(t) \no\\
 &=& -m\ddot{x}_0(t)x +c(t),
\ee
where $c(t)$ is an irrelevant function which is proportional to 
the identity operator.
We can conclude that this counterdiabatic potential works for 
arbitrary bound states $\varphi_n(x,t)$
in the present system with the potential (\ref{Utr}).

Second, we consider the case where the potential takes the form 
\be
 U(x,t)=\frac{1}{\xi^2(t)}U_0\left(\frac{x}{\xi(t)}\right).
\ee
A positive function $\xi(t)$ represents the dilatation.
With this potential, the probability density is given by 
\be
 \rho_n(x,t)=\frac{1}{\xi(t)}f_n\left(\frac{x}{\xi(t)}\right),
\ee
where $f$ is a positive function.
Then, the phase is given by 
\be
 \phi_n(x,t) &=& -\frac{m}{2}\frac{\dot{\xi}(t)}{\xi(t)}
 (x^2-{x^*}^2)
 \no\\ & & 
 +m\dot{\xi}(t)x^*
 f_n\left(\frac{x^*}{\xi(t)}\right)
 \int_{x^*/\xi(t)}^{x/\xi(t)}\frac{\diff z_1}{f_n\left(z_1\right)}.
 \label{phid0}
\ee
The last term is neglected if we set $f_n(x^*/\xi(t))=0$.
In that case, we obtain the corresponding counterdiabatic potential 
\be
 V_n(x,t) = -\frac{m}{2}\frac{\ddot{\xi}(t)}{\xi(t)}x^2 +c(t).
\ee
The last term represents an irrelevant shift.

The same systems are analyzed in Refs.~\cite{delCampo,dCB,DJdC}
as the scale-invariant driving.
A system in a moving harmonic potential is analyzed in Ref.~\cite{GM}.
Our results are consistent with theirs.
However, we note that 
the potential depends on the adiabatic state in principle.
The dependence can be included to an irrelevant term in principle 
but we can consider the state dependent potential 
by keeping the last term of Eqs.~(\ref{phix0}) and (\ref{phid0}).

Our formulation can also be applied to 
systems with the Lewis-Leach potential
which are generalizations of the above examples~\cite{LL}.
Fast-forwarding of the state was studied in Ref.~\cite{TMRM} 
and we can perform a similar analysis to that system.

As we mentioned in the Introduction, 
the idea of using the unitary transformation is the same as  
in the fast-forward method~\cite{MN1,MN2,MN3,TMRM,MR,Takahashi3}. 
However, instead of the time reparametrization, 
the acceleration of the adiabatic motion
is considered in Ref.~\cite{TMRM} and in our formulation.
Furthermore, we can consider a more generalized form of the unitary
transformation which cannot be interpreted as the addition of the phase
as in Eq.~(\ref{phase}). 
This transformation is demonstrated in the example of the next section.

\subsection{Generalization}
\label{gen}

We extend the method to systems in arbitrary dimensions.
We write the original adiabatic state as 
\be
 \psi_n(\bm{r},t) = \e^{-i\int_0^t \diff t'\,E_n(t')-i\phi_n^{(0)}(\bm{r},t)}
 \sqrt{\rho_n(\bm{r},t)}.
\ee
Introducing the new state 
$\tilde{\psi}_n(\bm{r},t)=\e^{-i\phi_n(\bm{r},t)}\psi_n(\bm{r},t)$, 
we obtain the continuity equation
\be
 && \frac{\partial\rho_n}{\partial t}
 =\frac{1}{m}\bm{\nabla}\cdot\left(
 \rho_n\bm{\nabla}\phi_n
 \right), \label{phixd}
\ee
and the potential 
\be
 V_n=\dot{\phi}_n
 -\frac{1}{2m}\left(\bm{\nabla}\phi_n\right)^2
 +\dot{\phi}_n^{(0)}
 -\frac{1}{m}\bm{\nabla}\phi_n^{(0)}\cdot\bm{\nabla}\phi_n. \label{Vxd}
\ee
We note that the original adiabatic state satisfies the relation
\be
 \bm{\nabla}\cdot \left(\rho_n\bm{\nabla}\phi_n^{(0)}\right)=0,
\ee
which means that the original current is in the form of 
$\bm{j}^{(0)}=\bm{\nabla}\times \bm{A}$.

Once we find the original wave function, 
we can calculate the potential by solving the above equations.
It is interesting to see that the problem of the level degeneracy 
does not exist in the present method.
It was a serious problem of the counterdiabatic driving where 
the counterdiabatic term (\ref{Hcd}) goes to infinity at the degenerate point.
Instead, the phase can be infinity when 
the probability density becomes zero at some point.
This problem was recognized in the fast-forward 
method~\cite{MN2,MN3,Takahashi3}.

The form of the counterdiabatic potential depends on 
the adiabatic state in principle.
In the following, we treat several examples 
to study typical properties of the potential.
 
\subsection{$1/r$ potential}
\label{r-1}

We study the potential for the hydrogen atom: 
\be
 U(\bm{r},t)=-\frac{1}{m\xi(t)|\bm{r}-\bm{r}_0(t)|}.
\ee
$\bm{r}_0(t)$ describes translation and $\xi(t)$ dilatation.
The wave function of the ground state is real and we have $\phi^{(0)}=0$.
The probability density is given by 
\be
 \rho(\bm{r},t)=\frac{1}{\pi\xi^3(t)}
 \exp\left(-\frac{2|\bm{r}-\bm{r}_0(t)|}{\xi(t)}\right).
\ee

We first consider the case with $t$-independent $\xi$.
Then, the phase $\phi$ becomes a function of $\phi(r(t))$
where $r(t)=|\bm{r}-\bm{r}_0(t)|$.
Solving the continuity equation, we obtain 
\be
 \frac{\partial\phi}{\partial r(t)}
 =m\dot{r}(t)\left(1+\frac{\xi}{r(t)}+\frac{\xi^2}{2r^2(t)}\right),
\ee
where we take the reference point $\bm{r}^*$ to be infinity.
Correspondingly, the potential is 
\be
 V &=& 
 m\ddot{r}(t)\left(r(t)+\xi\ln r(t)-\frac{\xi^2}{2r(t)}\right)
 \no\\ & &
 -\frac{m\dot{r}^2(t)}{2}\frac{\xi^2}{r^2(t)}
 \left(1+\frac{\xi}{2r(t)}\right)^2.
\ee
This result can be understood qualitatively.
We need a strong attraction at $\bm{r}=\bm{r}_0(t)$
to avoid deviation from the trapped state in a moving potential.

In the case of $\bm{r}_0(t)=0$,
$\phi$ is a function of $z(r,t)=r/\xi(t)$ and $t$.
We obtain 
\be
 \frac{\partial\phi}{\partial z}=-m\xi\dot{\xi}
 \left(z-\frac{1}{2}-\frac{1}{4z}\right),
\ee
and 
\be
 V &=& -\frac{m\xi\ddot{\xi}}{2}\left(
 z^2-z-\frac{1}{2}\ln z\right) 
 \no\\ & &
 -\frac{m\dot{\xi}^2}{2}\left(
 -z+\frac{1}{4}-\frac{1}{2}\ln z +\frac{1}{4z}+\frac{1}{16z^2}\right).
\ee

These results show that the counterdiabatic potential depends on the form of 
the ground-state wave function and takes a complicated form.
There is no useful property in three-dimensional systems 
as we found in the one-dimensional systems.

\section{Finite Hilbert space}
\label{discrete}

We apply the unitary deformation to systems 
in a finite-dimensional Hilbert space.

\subsection{General discussions}
\label{general}

We consider an $N$-dimensional system.
The adiabatic Hamiltonian is given by Eq.~(\ref{Had}).
The corresponding adiabatic state is written as 
\be
 |\psi_n(t)\rangle =\exp\left[-i\int_0^t \diff t'\,\left(
 E_n(t')-i\langle n(t')|\dot{n}(t')\rangle\right)
 \right]|n(t)\rangle. \no\\
\ee
We consider the unitary transformation 
\be
 \hat{U}_n(t) = \exp\left(-i\hat{\phi}^{(n)}(t)\right),
\ee
where $\hat{\phi}^{(n)}(t)$ is a diagonal operator in some basis 
written as 
\be
 \hat{\phi}^{(n)}(t) 
 = {\rm diag} (\phi_1^{(n)}, \phi_2^{(n)}, \ldots, \phi_N^{(n)}).
 \label{phidiag}
\ee
Our goal is to find the potential in a diagonal form: 
\be
 \hat{V}_n(t) 
 = {\rm diag} (v_1^{(n)}, v_2^{(n)}, \ldots, v_N^{(n)}). 
\ee

The unitary-transformed state satisfies the Schr\"odinger equation
\be
 i\frac{\diff}{\diff t}|\tilde{\psi}_n(t)\rangle 
 = \left(\hat{H}_{\rm ad}(t)+\hat{V}_n(t)\right)
 |\tilde{\psi}_n(t)\rangle, 
\ee
where $|\tilde{\psi}_n(t)\rangle=\hat{U}_n(t)|\psi_n(t)\rangle$.
This condition gives 
\be
 &&\sum_{a=1}^N(\dot{\phi}_a^{(n)}-v_a^{(n)})\hat{X}_a|n\rangle
 +i(1-|n\rangle\langle n|)|\dot{n}\rangle
 \no\\ &&
 +(E_n-\e^{i\hat{\phi}^{(n)}}\hat{H}_{\rm ad}\e^{-i\hat{\phi}^{(n)}})|n\rangle = 0.
\ee

We consider the case where the adiabatic Hamiltonian is a real symmetric matrix 
in the present representation.
Then, the eigenstate can be represented by a real vector
and we can decompose the above equation into the real and imaginary parts
as we did in the previous section.
We obtain 
\be
 &&(\dot{\phi}_a^{(n)}-v_a^{(n)}+E_n)\langle a|n\rangle
 \no\\ && 
 -\sum_{b=1}^N\langle a|\hat{H}_{\rm ad}|b\rangle
 \cos(\phi_a^{(n)}-\phi_b^{(n)})\langle b|n\rangle = 0,  
 \\
 && \langle a|\dot{n}\rangle
 -\sum_{b=1}^N\langle a|\hat{H}_{\rm ad}|b\rangle
 \sin(\phi_a^{(n)}-\phi_b^{(n)})\langle b|n\rangle = 0, 
\ee
where $\langle a|n\rangle$ denotes the $a$th component 
of the vector $|n\rangle$
and $\langle a|\hat{H}|b\rangle$ denotes 
the $(a,b)$ component of the matrix $\hat{H}$.
The label $a$ (and $b$) takes an integer between 1 and $N$.
We note that $\langle n|\dot{n}\rangle=0$.
These equations represent the Hamilton-Jacobi equation
and the continuity equation, respectively.
Since we consider discrete systems,
the equations are not written by differential ones.
We can obtain the phase $\phi_a^{(n)}$ from the second equation 
and the potential $v_a^{(n)}$ from the first.

Before considering examples, we study the state dependence of the potential.
We assume that $\phi_a^{(n)}$ and $v_a^{(n)}$ are independent of $n$.
Then, using the completeness relation, we obtain 
\be
 &&(\dot{\phi}_a-v_a)\delta_{ab}
 +\langle a|\hat{H}_{\rm ad}|b\rangle
 (1-\cos(\phi_a-\phi_b)) = 0, \label{udd1}
 \\
 && -i\langle a|\hat{H}_{\rm cd}(t)|b\rangle
 -\langle a|\hat{H}_{\rm ad}|b\rangle
 \sin(\phi_a-\phi_b) = 0, \label{udd2}
\ee
where the counterdiabatic Hamiltonian is written as 
\be
 \hat{H}_{\rm cd} = i\sum_n \left(1-|n\rangle\langle n|\right)
 |\dot{n}\rangle\langle n|
 = i\sum_n  |\dot{n}\rangle\langle n|.
\ee
These equations give 
$v_a=\dot{\phi}_a$ and $\langle a|\hat{H}_{\rm cd}|b\rangle=0$.
Thus, the present assumption only describes 
a trivial situation where the counterdiabatic Hamiltonian is zero.
We conclude that the potential $\hat{V}_n(t)$ crucially depends on 
the state $n$.
However, we can expect that the dependence appears only 
at the irrelevant energy shift for simple systems.
We study this property in the following simple example.

\subsection{Two-level systems}
\label{two}

As an example, we study the two-level systems described by the Hamiltonian 
\be
 \hat{H}_{\rm ad}(t) &=& \frac{h(t)}{2}\bmat{cc}
 \cos\theta(t) & \sin\theta(t) \\
 \sin\theta(t) & -\cos\theta(t) \emat. \label{Hxy}
\ee
The eigenvalues of $\hat{H}_{\rm ad}(t)$ are given by $\pm h(t)/2$.
Each adiabatic state is given, respectively, by 
\be
 && |\psi_1(t)\rangle = \e^{-i\int^t_0 \diff t'\,h(t')/2}\bmat{c} 
 \cos\frac{\theta(t)}{2} \\
 \sin\frac{\theta(t)}{2}\emat,  \label{psiadp}\\
 && |\psi_2(t)\rangle = \e^{i\int^t_0 \diff t'\,h(t')/2}\bmat{c} 
 -\sin\frac{\theta(t)}{2} \\
 \cos\frac{\theta(t)}{2}\emat. 
\ee

Equation (\ref{udd2}) is written as
\be
 \dot{\theta}+h\sin\theta\sin (\phi_1^{(n)}-\phi_2^{(n)})=0,
 \label{phid}
\ee
which means that the difference of the phases $\phi_1^{(n)}-\phi_2^{(n)}$
is independent of $n$.
The same is true for the potential and we have from Eq.~(\ref{udd1})
\be
 v_1^{(n)}-v_2^{(n)}=
 \dot{\phi}_1^{(n)}-\dot{\phi}_2^{(n)}
 -h(1-\cos(\phi_1^{(n)}-\phi_2^{(n)}))\cos\theta.
 \no\\
 \label{vd}
\ee
We obtain the form of the potential 
\be
 \hat{V}_n(t)=\frac{1}{2}v(t)\hat{\sigma}^z +\frac{1}{2}(v_1^{(n)}(t)+v_2^{(n)}(t)),
\ee
where $v(t)=v_1^{(n)}(t)-v_2^{(n)}(t)$.
By applying this potential, we can realize the time evolution of 
the state along an adiabatic passage.
In the two-dimensional case, 
the state dependence of the potential only appears in the energy shift.

We note that the adiabatic passage is different from
the original one.
The Bloch vector of the unitary-transformed state is not 
in the $zx$ plane.
The initial and final states of $|\tilde{\psi}(t)\rangle$ 
should be the same as those of the original state $|\psi(t)\rangle$.
We demand that the potential goes to zero at initial and final times.
This is achieved when $\phi_1-\phi_2=0$ and $\dot{\phi}_1-\dot{\phi}_2=0$ 
at those times.
From the condition (\ref{phid}), 
we obtain $\dot{\theta}=0$ and $\ddot{\theta}=0$,
which means that the potential should be turned on and off very slowly.
It would be interesting to understand the relation with 
the analysis of Ref.~\cite{Morita} where 
the same condition is obtained for improving the adiabatic approximation.

As a simple example, we consider 
\be
 && \hat{H}_{\rm ad}(t)
 =\frac{1}{2}\bm{h}(t)\cdot\hat{\bm{\sigma}}
 =\frac{1}{2}h(t)\bm{n}(t)\cdot\hat{\bm{\sigma}}, \\
 &&\bm{h}(t)=\bmat{c} \Gamma \\ 0 \\ h_z(t)\emat.
\ee
$\theta(t)$ is represented as 
\be
 \cos\theta(t)=\frac{h_z(t)}{h(t)},
\ee
where $h(t)=\sqrt{\Gamma^2+h_z^2(t)}$.
The Bloch vector $\bm{n}(t)$ is parametrized as 
\be
 \bm{n}(t)=(\sin\theta(t), 0, \cos\theta(t)).
\ee
We impose the condition 
$\theta(0)=\pi/2$ and $\theta(\infty)=0$, which means 
that the Bloch vector moves from $(1,0,0)$ to $(0,0,1)$
in the $zx$ plane.

\begin{figure}[tb]
\includegraphics[width=1.\columnwidth]{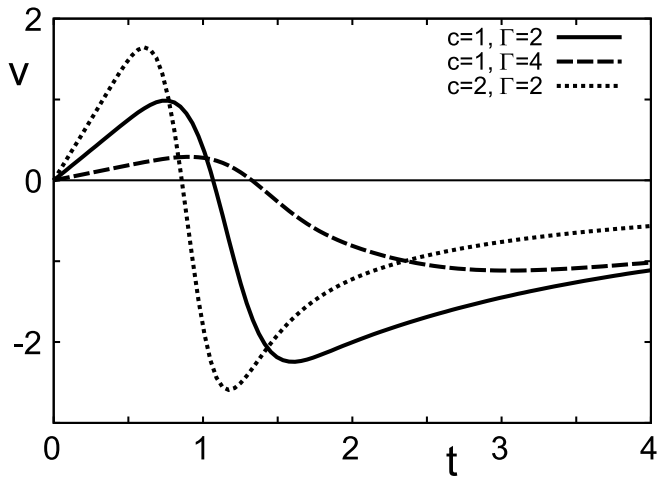}
\caption{Potential $v(t)=v_1(t)-v_2(t)$ in Eq.~(\ref{vd}).
The phase difference $\phi_1-\phi_2$ is determined from 
Eqs.~(\ref{phi2}) and (\ref{ct3}).
}
\label{fig1-v}
\end{figure}

\begin{figure}[tb]
\includegraphics[width=1.\columnwidth]{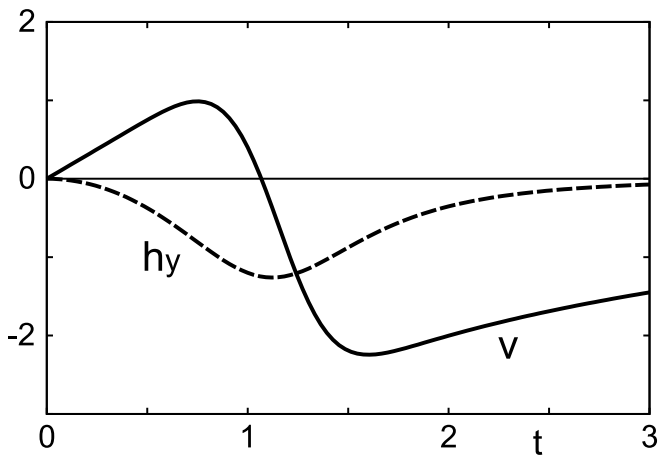}
\caption{$v(t)$ at $c=1$ and $\Gamma=2$.
$h_y=\dot{\theta}$ represents the counterdiabatic field in Eq.~(\ref{hy}).}
\label{fig2-hy}
\end{figure}

\begin{figure}[tb]
\includegraphics[width=1.\columnwidth]{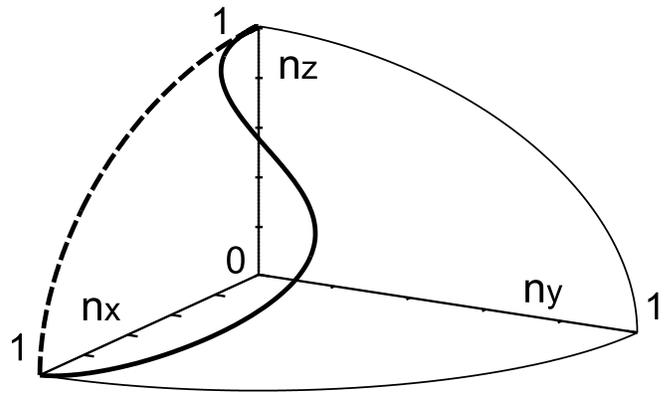}
\caption{Bloch vectors $\bm{n}(t)$ 
before the deformation (dashed line) 
and $\tilde{\bm{n}}(t)$ after the deformation (bold).
$\bm{n}(0)=\tilde{\bm{n}}(0)=(1,0,0)$
and $\bm{n}(\infty)=\tilde{\bm{n}}(\infty)=(0,0,1)$.
We set $c=1$ and $\Gamma=2$.}
\label{fig3-bloch}
\end{figure}

In the unitary-deformed evolution, 
the Bloch vector is not in the plane.
Equation (\ref{phid}) is written as 
\be
 && \sin(\phi_1(t)-\phi_2(t))=\frac{\dot{h}_z(t)}{h^2(t)}.
 \label{phi2}
\ee
To satisfy the condition that $\phi_1-\phi_2$ and 
$\dot{\phi}_1-\dot{\phi}_2$ go to zero
at $t=0$ and $\infty$, 
we take the magnetic field in $z$ direction as
\be
 h_z(t)=ct^3, \label{ct3}
\ee
where $c$ is a constant.
The potential $v(t)$ obtained from Eq.~(\ref{vd}) 
is plotted in Fig.~\ref{fig1-v}.
For comparison, we plot 
the potential $v$ and the counterdiabatic field 
$h_y=\dot{\theta}$ in Fig.~\ref{fig2-hy}.
We note that the counterdiabatic field is given by 
\be
 &&\hat{H}_{\rm cd}(t) 
 = \frac{1}{2}(\bm{n}(t)\times\dot{\bm{n}}(t))\cdot\bm{\sigma}
 = \frac{1}{2}\dot{\theta}(t)\sigma_y. \label{hy}
\ee
To see that the adiabatic passage of the unitary-deformed 
state is different from that of the original state, 
we plot the original Bloch vector $\bm{n}(t)$ and 
the vector after the deformation 
\be
 && \tilde{\bm{n}}(t)=(\sin\theta(t)\cos\phi(t), 
 \sin\theta(t)\sin\phi(t), \cos\theta(t)),
\ee
where $\phi(t)=\phi_1(t)-\phi_2(t)$.
They are plotted in Fig.~\ref{fig3-bloch}.

\subsection{Choice of the unitary transformation}

Up to this point, we used diagonal operators
for the unitary transformation 
as we see in Eq.~(\ref{phidiag}).
This choice is not necessary and we can consider a different type 
of the unitary transformation.
In two-level systems, the most general form of the unitary transformation
reads 
\be
 \hat{U}(t)=\exp\left(
 -i\phi_0(t)-i\phi(t)\bm{\sigma}\cdot\bm{n}(t)\right),
\ee
where $\bm{n}(t)$ represents a unit vector.
If we take for example
\be
 \bm{n}=\bmat{c} \sin\varphi \\ 0 \\ \cos\varphi \emat,
\ee
where $\varphi$ is independent of $t$, we obtain 
\be
 && \dot{\theta}=\left[v\sin\varphi-h\sin(\theta-\varphi)\right]\sin 2\phi, \\
 && \frac{v}{2}\left[
  (\cos^2\phi+\sin^2\phi\cos2\varphi)\sin\theta
 -\sin^2\phi\sin 2\varphi\cos\theta
 \right] \no\\
 && =\dot{\phi}\sin(\theta-\varphi)
 -\frac{h}{2}\sin 2(\theta-\varphi)\sin^2\phi.
\ee
For a given $\varphi$, 
$\phi$ and $v$ are obtained by solving these equations.
We note that the result at $\varphi=0$ corresponds to that in the previous 
subsection.
A different choice of $\varphi$ gives a different adiabatic passage 
and the control potential is changed accordingly.

Thus, we can consider many possible forms of the unitary transformation.
In the present simple example, this is an unnecessary prescription but 
in principle this can be useful when 
the standard deformation does not work.

\section{Quantum brachistochrone equation}
\label{qb}

In the previous section, we have formulated the unitary deformation
of the counterdiabatic driving.
Here we formulate the problem using the quantum brachistochrone equation.
This formulation allows us to understand 
what quantity characterizes the deformed system.

\subsection{Formulation of the problem}

We consider the Schr\"odinger equation in Eq.~(\ref{Sch})
with $\hat{H}(t)=\hat{H}_{\rm ad}(t)+\hat{H}_{\rm cd}(t)$.
It is deformed as 
\be
 i\frac{\diff}{\diff t}|\tilde{\psi}(t)\rangle 
 =(\hat{H}_{\rm ad}(t)+\hat{V}(t))|\tilde{\psi}(t)\rangle.
\ee
The potential $\hat{V}(t)$ is diagonal in a specific basis.
In the $N$-dimensional Hilbert space,
the number of independent traceless diagonal operators are $N-1$.
Their basis operators commute with each other:
\be
 [\hat{X}_a,\hat{X}_b]=0 \quad  (a,b=1,2,\ldots,N-1).
\ee
They satisfy the orthonormal relations
\be
 \Tr \hat{X}_a\hat{X}_b = \delta_{ab}.
\ee
Using these operators, we can generally write 
\be
 \hat{V}(t)=\sum_{a=1}^{N-1} v_a(t)\hat{X}_a,
\ee
where
\be
 v_a(t) = \Tr\hat{V}(t)\hat{X}_a.
\ee
The original state $|\psi(t)\rangle$ 
is changed by the unitary deformation 
to a deformed state $|\tilde{\psi}(t)\rangle$.
They are related by the constraints 
\be
 |\langle\sigma|\psi(t)\rangle|
 =|\langle\sigma|\tilde{\psi}(t)\rangle|,
\ee
where $|\sigma\rangle$ with $\sigma=1,2,\ldots,N$
represents a diagonal basis.
This condition implies that the deformed state is written as 
\be
 |\tilde{\psi}(t)\rangle = \hat{U}(t)|\psi(t)\rangle,
\ee
where the unitary operator is written in terms of 
diagonal operators 
\be
 \hat{U}(t)=\exp\left(-i\phi_0(t)-i\sum_{a=1}^{N-1}\phi_a(t)\hat{X}_a\right).
\ee
Using this setting we consider the optimization 
of the potential.

\subsection{Action}

The optimization problem is formulated by defining the action 
to be minimized.
It consists of four parts: 
\be
 S = \int \diff t\,(L_{\rm T}+L_{\rm S}+L_{\rm C1}+L_{\rm C2}).
\ee
The first term represents the time duration \cite{CHKO} 
\be
 L_{\rm T}=\sqrt{\frac{\langle\dot{\tilde{\psi}}|(1-|\tilde{\psi}\rangle
 \langle\tilde{\psi})|\dot{\tilde{\psi}}\rangle}{\langle\tilde{\psi}|
 (\hat{H}_{\rm ad}+\hat{V})^2|\tilde{\psi}\rangle
 -\langle\tilde{\psi}|\hat{H}_{\rm ad}+\hat{V}|\tilde{\psi}\rangle^2}},
\ee
which means 
the Fubini-Study distance divided by the velocity based on 
the Anandan-Aharonov relation.
The other terms represent constraints.
The second term is for the Schr\"odinger equation:
\be
 L_{\rm S} = \langle\phi|\left(i\frac{\diff}{\diff t}
 -(\hat{H}_{\rm ad}+\hat{V})\right)|\tilde{\psi}\rangle +({\rm h.c.}),
\ee
where $|\phi\rangle=|\phi(t)\rangle$ 
plays the role of a multiplier function.
The third term denotes constraints for the potential:
\be
 L_{\rm C1}=
 \lambda_0\Tr\hat{V}
 +\sum_{b=1}^{N^2-N}\lambda_b\Tr\hat{V}\hat{Y}_b.
\ee
where $\{\hat{Y}_b\}_{b=1,\ldots,N^2-N}$ are traceless off-diagonal operators.
They are the complements of diagonal operators 
$\{\hat{X}_a\}_{a=1,\ldots,N-1}$ and
satisfy 
\be
 \Tr \hat{X}_a\hat{Y}_b = 0, 
\ee
for arbitrary $a$ and $b$.
$L_{\rm C1}$ makes the trace and the off-diagonal elements of the potential zero.
The last term of the action represents constraints for 
the state $|\tilde{\psi}\rangle$.
We write
\be
 L_{\rm C2}=\sum_{\sigma=1}^N\lambda_\sigma\Bigl(|\langle\sigma|\tilde{\psi}\rangle|^2
 -|\langle\sigma|\psi\rangle|^2
 \Bigr). \label{Lc2}
\ee

We note that the last term $L_{\rm C2}$ 
is not present in the original formulation \cite{CHKO}.
The constraint for the state has not been studied before.

\subsection{Quantum brachistochrone equation}

Now that the action is defined, 
we can derive the quantum brachistochrone equation 
from the extremized condition.
The extremization is performed on the potential and the state.
We have 
\be
 0 &=&  \frac{\delta S}{\delta \langle\tilde{\psi}|} \no\\
 &=&  
 i\frac{\diff}{\diff t}\left[\frac{\hat{H}_{\rm ad}+\hat{V}
 -\langle\tilde{\psi}|(\hat{H}_{\rm ad}+\hat{V})|\tilde{\psi}\rangle}
 {2\Delta E^2}\right]|\tilde{\psi}\rangle\no\\
 && +\left[i\frac{\diff}{\diff t}-(\hat{H}_{\rm ad}+\hat{V})\right]|\phi\rangle
 +\sum_\sigma\lambda_\sigma|\sigma\rangle\langle\sigma|\tilde{\psi}\rangle,
 \label{var1}\\
 0 &=& \frac{\delta S}{\delta\hat{V}} \no\\
 &=& -\frac{1}{2\Delta E^2}\Bigl[(\hat{H}_{\rm ad}+\hat{V})
 \hat{P}
 +\hat{P}(\hat{H}_{\rm ad}+\hat{V})
 \no\\
 && 
 -2\langle\tilde{\psi}|(\hat{H}_{\rm ad}+\hat{V})|\tilde{\psi}\rangle\hat{P}
 \Bigr]
 -\left(|\tilde{\psi}\rangle\langle\phi|+|\phi\rangle\langle\tilde{\psi}|\right)
 +F. \no\\
 \label{var2}
\ee
We use the following notations: 
\be
 && \hat{P}=|\tilde{\psi}\rangle\langle\tilde{\psi}|, \\
 &&\Delta E^2=
 \langle\tilde{\psi}|(\hat{H}_{\rm ad}+\hat{V})^2|\tilde{\psi}\rangle
 -\langle\tilde{\psi}|(\hat{H}_{\rm ad}+\hat{V})|\tilde{\psi}\rangle^2, \\
 && \hat{F} = \sum_{b=0}^{N^2-N} \lambda_b \hat{Y}_b.
\ee
These are time-dependent functions.

Equation (\ref{var1}) is for $|\phi\rangle$ and is written as 
\be
 i\frac{\diff}{\diff t}|\phi\rangle
 &=& (\hat{H}_{\rm ad}+\hat{V})|\phi\rangle
 -i\frac{\diff \hat{D}}{\diff t}
 |\tilde{\psi}\rangle
 -\sum_\sigma\lambda_\sigma|\sigma\rangle\langle\sigma|\tilde{\psi}\rangle,
 \no\\ 
 \\
 \hat{D} &=&
 \frac{\hat{H}_{\rm ad}+\hat{V}
 -\langle\tilde{\psi}|(\hat{H}_{\rm ad}+\hat{V})|\tilde{\psi}\rangle}
 {2\Delta E^2}.
\ee
The solution is written in the form 
\be
 |\phi(t)\rangle &=& \hat{T}(t)\biggl[|\phi(0)\rangle
 -(\hat{T}^\dag(t) \hat{D}(t)\hat{T}(t)-\hat{D}(0))|\tilde{\psi}(0)\rangle
 \no\\
 && +i\sum_\sigma\int^t_0 \diff t'\,
 \lambda_\sigma(t')\hat{T}^\dag(t')|\sigma\rangle\langle\sigma|
 \hat{T}(t')|\tilde{\psi}(0)\rangle\biggr], \no\\
\ee
where $\hat{T}(t)$ is the time-evolution operator 
of the Hamiltonian $\hat{H}_{\rm ad}(t)+\hat{V}(t)$
with the initial condition $\hat{T}(0)=1$.
Inserting this expression into the second equation (\ref{var2}), we obtain
\be
 \hat{F}(t) &=& 
 |\tilde{\psi}(t)\rangle\langle\phi(t)|
 +|\phi(t)\rangle\langle\tilde{\psi}(t)| \no\\
 & & 
 +\hat{D}(t)|\tilde{\psi}(t)\rangle\langle\tilde{\psi}(t)|
 +|\tilde{\psi}(t)\rangle\langle\tilde{\psi}(t)|\hat{D}(t) \no\\
 &=& \hat{T}(t)
 \Bigl\{
 |\tilde{\phi}(0)\rangle\langle\tilde{\psi}(0)|
 +|\tilde{\psi}(0)\rangle\langle\tilde{\phi}(0)|\no\\
 & & 
 +i\left[\hat{Z}(t), \hat{P}(0)\right]
 \Bigr\}
 \hat{T}^\dag(t),
\ee
where 
\be
 && |\tilde{\phi}(0)\rangle
 =|\phi(0)\rangle +\hat{D}(0)|\tilde{\psi}(0)\rangle,
 \\
 &&\hat{Z}(t)=\sum_\sigma\int^t_0 \diff t'\,
 \lambda_\sigma(t')\hat{T}^\dag(t')|\sigma\rangle\langle\sigma|
 \hat{T}(t'). \label{Zt}
\ee
The operator $\hat{F}(t)$ satisfies the differential equation 
\be
 i\frac{\diff\hat{F}(t)}{\diff t}
 &=& \left[\hat{H}_{\rm ad}(t)+\hat{V}(t), \hat{F}(t)\right]
 \no\\
 & & -\sum_\sigma\lambda_\sigma(t)\left[
 |\sigma\rangle\langle\sigma|,
 \hat{P}(t)\right].
\ee
If the second term of the right-hand side is absent,
this equation indicates that 
$\hat{F}(t)$ is a dynamical invariant quantity~\cite{LR}.
The presence of the state-constraint term 
changes this standard interpretation.

At $t=0$, $\hat{F}$ is written as
\be
 \hat{F}(0)=|\tilde{\phi}(0)\rangle\langle\tilde{\psi}(0)|
 +|\tilde{\psi}(0)\rangle\langle\tilde{\phi}(0)|,
\ee
where $|\tilde{\psi}(0)\rangle$ is an arbitrary vector.
We assume that the initial state is one of 
the eigenstates of $\hat{H}_{\rm ad}(0)$ and
the potential is absent at $t=0$.
Thus, we choose
\be
 \hat{F}(0)=|\tilde{\psi}(0)\rangle\langle\tilde{\psi}(0)|.
\ee
Then, 
\be
 \hat{F}(t)=\hat{P}(t)+i\hat{T}(t)[\hat{Z}(t),\hat{P}(0)]\hat{T}^\dag(t).
\ee
This equation shows that the operator defined as
\be
 \hat{I}(t)= \hat{F}(t)-i\hat{T}(t)[\hat{Z}(t),\hat{P}(0)]\hat{T}^\dag(t)
 \label{It}
\ee
represents the dynamical invariant in the unitary-deformed evolution.
This operator satisfies the equation for the invariant
\be
 i\frac{\diff\hat{I}(t)}{\diff t}
 =\left[\hat{H}_{\rm ad}(t)+\hat{V}(t), \hat{I}(t)\right].
\ee

The quantum brachistochrone equation shows 
that the unitary-deformed system is characterized by 
a dynamical invariant.
It is different from the invariant of the original state 
before the deformation.
The formulation using the invariant was done in 
Ref.~\cite{DJdC} for scale-invariant systems.
Our result implies that the method of the unitary deformation 
is equivalent with the invariant-based method and 
can be applied to general systems. 

\subsection{Two-level systems}

As an example, we consider the two-level system treated in 
Sec.~\ref{two}.
The adiabatic Hamiltonian is given in Eq.~(\ref{Hxy}) and 
we consider one of the adiabatic states in Eq.~(\ref{psiadp}).
The constraints for the Hamiltonian are represented as 
\be
 \hat{F}(t)=\lambda_0+\lambda_3(t)\sigma^z+\lambda_1(t)\sigma^x,
\ee
In the counterdiabatic driving, this operator represents 
the invariant as 
\be
 \hat{F}(t)=|\psi_1(t)\rangle\langle\psi_1(t)|
 =\frac{1}{2}\bmat{cc} 1+\cos\theta(t) & \sin\theta(t) \\
 \sin\theta(t) &  1-\cos\theta(t) \emat, \no\\
\ee
and the counterdiabatic field is given by Eq.~(\ref{hy}).
In the presence of the additional constraint for the state (\ref{Lc2}), 
the equation for $\hat{F}(t)$ is written as
\be
 i\frac{\diff\hat{F}(t)}{\diff t}
 &=& \left[\hat{H}_{\rm ad}(t)+\hat{V}(t), \hat{F}(t)\right]
 -\frac{\lambda_z(t)}{2}\left[
 \sigma^z, \hat{P}(t)\right]. \no\\
\ee 
The last term affects the $\sigma^x$ and $\sigma^y$ terms in $\hat{F}(t)$
and changes the optimal path as we see in Fig.~\ref{fig3-bloch}. 
The invariant in Eq.~(\ref{It}) is given by 
\be
 \hat{I}(t) &=& \lambda_0+\lambda_3(t)\sigma_z 
 +\left(\lambda_1(t)+\tilde{\lambda}_z(t)\sin\phi(t)\right)\sigma^x \no\\
 && +\tilde{\lambda}_z(t)\cos\phi(t)\sigma^y, 
 \label{It2} 
\ee
 where
\be
 \tilde{\lambda}_z(t)=\int_0^t\diff t\,\lambda_z(t')\sin\theta(t').
\ee
The coefficient $\tilde{\lambda}_z(t)$ depends on the history of
the time evolution, which means that the optimized path cannot be  
determined from the geometry of the original adiabatic state.
Imposing the condition that Eq.~(\ref{It2}) represents the invariant, 
we can find the explicit form of the potential $\hat{V}(t)$.
We do not explicitly perform such a calculation here since 
it is not a convenient method for finding the potential.
The present formulation is conceptually important and 
will be useful to understand the method from a general principle.

\section{Summary}
\label{sum}

We have formulated the method of unitary deformation.
The counterdiabatic Hamiltonian which supports 
the adiabatic state evolution is represented by a local potential.
The main results are presented in Eqs.~(\ref{phixd}) and (\ref{Vxd}) 
for potential systems 
and Eqs.~(\ref{phid}) and (\ref{vd}) for discrete systems.
The advantage of this method is that 
the counterdiabatic potential is constructed 
from the single adiabatic state.
It is not necessary to know all eigenstates of the adiabatic Hamiltonian.
The form of the potential is written 
immediately once we specify the adiabatic state.
Instead of this property, the form of the potential is dependent on 
the state to accelerate.
However, for several cases such as two-level systems and 
one-dimensional systems with the scale invariance,
the dependence can be reduced to the irrelevant energy shift.

The potential is not unique in principle.
For example, by keeping the last term in Eq.~(\ref{phix0})
or by extending the unitary transformation
to a more general one, 
we can find a different form of the potential.
Due to this arbitrariness, we have 
many possibilities to find a more convenient form of the potential.

From the analysis of the quantum brachistochrone equation, 
we find that our method can be formulated by using the dynamical invariant
as was done in other acceleration methods~\cite{STA}.
Our formulation may not be useful to find the explicit form of the invariant.
Rather, it is important to notice 
that we can formulate the methods in a unified way.
We expect that the present analysis enhances the applicability of the method
of shortcuts to adiabaticity.

\section*{Acknowledgments}

The author is grateful to S.~Masuda for stimulating discussions.
This work was supported by Japan Society for the Promotion of Science 
KAKENHI Grant No. 26400385.


\end{document}